\scrollmode

\documentclass[12pt,twoside,a4paper]{article}
\usepackage[dvips]{epsfig}

\def\Journal#1#2#3#4{{#1}#2 (#4) #3}
\def\NPB{{Nucl.~Phys.} B}
\def\PLB{{Phys.~Lett.}  B}

\def\PRD{{Phys.~Rev.} D}
\def\ZPC{{Z.~Phys.} C}
\def\EPJC{{Eur.~Phys.~J.} C}
\newcommand{\g}{\gamma}
\newcommand{\GeV}{\mbox{GeV}}
\newcommand{\err}{\mbox{$\pm$}}
\newcommand{\jv}{{\tt JetViP}}

\newcommand{\simgt}{\,\rlap{\lower 3.5 pt \hbox{$\mathchar \sim$}} \raise 1pt
 \hbox {$>$}\,}
\newcommand{\simlt}{\,\rlap{\lower 3.5 pt \hbox{$\mathchar \sim$}} \raise 1pt
 \hbox {$<$}\,}

\begin{document}
\hfill MPI-PhT 99-054
\vspace*{10mm}
\begin{center}  \begin{Large} \begin{bf}
{\tt JetViP 2.1}: the hbook version
\\ \end{bf}  \end{Large} \vspace*{5mm}
  \begin{large}
B.~P\"otter \\
  \end{large}
Max-Planck-Institut f\"ur Physik, M\"unchen, Germany\\ 
\end{center}
\begin{quotation}
\noindent
{\bf Abstract:}
We present an update of the \jv\ 1.1 program for performing fixed NLO
calculations in jet production including direct and resolved
components in a continuous range of photon virtuality $Q^2$.  
The new version allows to access the full event record on the parton
level. The program is set up such that {\tt hbook} can be used to fill
histograms. The phase space generator has been optimized and the
azimuthal dependence of the cross sections is taken into account in
LO. We comment on recent comparisons between various NLO
programs for jet production at HERA. We demonstrate that the 
$\sum E_T$ cut for dijet cross sections is not infrared safe.
\end{quotation}

\section{Introduction}

Several fixed order programs are on the market now for calculating jet
cross sections for non-zero photon virtuality $Q^2$ at HERA, as there
are {\tt DISENT} \cite{1}, {\tt DISASTER++} \cite{2}, {\tt MEPJET}
\cite{3} and \jv\ \cite{4}, which is based on the calculations in
\cite{5,6}. The first two programs are based on the subtraction method,
whereas the latter two employ the phase-space-slicing method. 
In the photoproduction limit $Q^2\to 0$, the deep-inelastic scattering
(DIS) matrix elements with a direct coupling of the photon to the partons
from the proton show an additional initial state singularity on the
photon leg, which has to be subtracted and absorbed in the photon
structure function. Furthermore, one has to calculate a so-called resolved
component, where the photon serves as a source of partons. These direct and
resolved photon-proton scattering processes have been calculated by
three groups in NLO QCD \cite{Frixione:1997ks,Harris:1997hz,%
Klasen:1996it}. The calculation of \cite{Frixione:1997ks} uses the
subtraction method, whereas the other two groups \cite{Harris:1997hz,%
Klasen:1996it} use the phase-space slicing method. One of
the main features of {\tt JetViP} is the possibility to include a
resolved virtual photon component in NLO. In this way the photoproduction
limit can be taken, which allows to cover the full range in $Q^2$
accessible at HERA.

The phase-space generator of \jv\ was originally built around the
$R_{sep}$-modified cone-algorithm. Though the phase-space could be
optimized for this specific case, the program was obviously limited in its
applicability and furthermore it was rather inconvenient
to calculate spectra of variables other than the jet transverse energy
and rapidity. Therefore, in the new version partons rather than jets
are accessible and an event record is provided in a user-routine, i.e.,
the four-vectors of all particles involved in the process are
available together with the basic kinematical variables such as
$Q^2,x_{Bj},y,\ldots$ and the weight from the matrix elements
convoluted with the parton distribution functions. The analysis of the
events is then left entirely to the user routine. 

Two further important improvements have been achieved. First, the
phase space generator has been rewritten, as to generate events in
$(y,Q^2)$-space, which greatly improves the statistics. 
Second, the azimuthal dependence of the matrix elements, which has some
effect for cuts in the laboratory-frame, have been implemented in
LO. As we will demonstrate, this is sufficient for small $Q^2$ and when
the NLO corrections are not too large. 

Recently, a comparison of the above mentioned NLO programs for DIS has 
become available \cite{comp}. We have recalculated the cross sections
produced by \jv\ presented in \cite{comp} with the new version. As
will become clear, differences in the cross sections produced by the
programs are due either to statistical problems or to infrared (IR)
sensitive cuts on dijet variables. Similarly, a comparison of
photoproduction calculations has been performed \cite{hkv}. We
compare photoproduction cross sections from \jv\ to those from one of
the programs used in \cite{hkv} and briefly comment on the implications
for the DIS region.

Before discussing these more physical topics, we describe the new
user-routine with common-blocks and the new steering file. Further
details of the \jv\ program can be found \cite{4}.

\section{Program changes}

\subsection{{\tt user} routine}

The events generated by \jv\ have to be analyzed in a routine called
{\tt user}.

{\tt subroutine user(iflag,weight,ifill) \\
 real*8 vgweight,weight,Qs,shad,Ws,xbj,y,xa,xb,plab,pcms\\
 integer iflag,ifill\\
 common/parton/plab(7,7),pcms(7,7)\\
 common/kin/Qs,shad,Ws,xbj,y,xa,xb \\
 common/vgplot/vgweight}

\begin{itemize}
\item {\tt iflag}: the routine {\tt user} is called once at the very
   beginning of \jv\ for initialization with {\tt iflag}=1 and {\tt
   weight}=0d0. After this initialization, events are generated with
   {\tt iflag}=2. At the 
   very end {\tt user} is called for termination with {\tt iflag}=3 
   and {\tt weight}=0d0. 
\item {\tt weight}: weight of the event, i.e., the
   matrix elements convoluted with the PDF's from the proton and/or
   photon, together with running $\alpha_{QED}$ and $\alpha_s$ (in
   pb). The entry for the histogram has to be further
   multiplied with the weight from {\tt VEGAS}, which is {\tt vgweight}. 
   The histogram entry therefore has to be {\tt vgweight*weight}.
\item {\tt ifill}: for {\tt ifill=0} the phase space is scanned by 
  {\tt VEGAS} with low statistics. In this stage the histograms 
  {\it must not} be filled. For {\tt ifill=1} the histograms can be filled. 
\item {\tt plab(7,7)}: matrix containing the four-vectors,
$E_T,\eta$ and $\phi$ for 7 different particles in the laboratory
frame. The $z$-direction is given by the incoming proton or, for the
$\gamma^*\gamma$ mode, by the real photon. 1st entry: particle. 2nd
entry: variable.
\[  \left( \begin{array}{c} 
 \mbox{\tt plab(1,i)} \\
 \mbox{\tt plab(2,i)} \\
 \mbox{\tt plab(3,i)} \\
 \mbox{\tt plab(4,i)} \\
 \mbox{\tt plab(5,i)} \\
 \mbox{\tt plab(6,i)} \\
 \mbox{\tt plab(7,i)}
 \end{array} \right) =  \left( \begin{array}{l} 
 \mbox{\tt incoming electron} \\ 
 \mbox{\tt outgoing electron} \\ 
 \mbox{\tt incoming proton} \\ 
 \mbox{\tt incoming photon} \\ 
 \mbox{\tt outgoing parton 1}\\
 \mbox{\tt outgoing parton 2}\\ 
 \mbox{\tt outgoing parton 3}
 \end{array} \right) \quad , \quad 
 \left( \begin{array}{c} 
 \mbox{\tt plab(j,1)} \\
 \mbox{\tt plab(j,2)} \\
 \mbox{\tt plab(j,3)} \\
 \mbox{\tt plab(j,4)} \\
 \mbox{\tt plab(j,5)} \\
 \mbox{\tt plab(j,6)} \\
 \mbox{\tt plab(j,7)}
 \end{array} \right) = 
 \left( \begin{array}{c} E_T \\ \eta \\ \phi \\ 
   p_x \\ p_y \\ p_z \\ E \end{array} \right)  \] 
The outgoing partons are not ordered in $E_T$. The incoming photon
vector, i.e., {\tt plab(4,i)}, is set equal to zero.
\item {\tt pcms(7,7)}: matrix containing the four-vectors,
$E_T,\eta$ and $\phi$ for 7 different particles in the hadronic, i.e.,
$\gamma^*P$ cms frame. In the $\gamma^*\gamma$ mode the $\gamma^*e$ cms
frame is chosen. The $z$-direction is given by the incoming
proton/real photon. Same conventions as for {\tt plab(7,7)}. The incoming and
outgoing electron vectors, i.e., {\tt pcms(1,i)} and {\tt pcms(2,i)},
are set equal to zero.
\item {\tt Qs,shad,Ws}: photon virtuality $Q^2=-q^2$, total cms
energy $s_H=4E_aE_b$ and hadronic cms energy $W^2=(P+q)^2$.
\item {\tt xbj,y,xa,xb} $\in \{0,1\}$: Bjorken-$x$ variable
$x_{bj}$, $y$-variable, momentum fraction of the parton in resolved
photon $x_a$ and momentum fraction of the parton in proton $x_b$. 
\end{itemize}

\subsection{Steering file}

An example of the steering file is given in appendix B.

\subsubsection*{Contribution}

\begin{itemize}
\item {\tt iproc}: integer $\in \{ 1,2\}$. Type of process.
  $1=eP$, $2=e^+e^-$ ($\gamma^*\gamma$-mode)
\item {\tt isdr}: integer $\in \{ 1,2,3,4\}$. Selection of
  the D, SR, SR$^*$ or DR contribution. $1=$D; $2=$SR;
  $3=$SR$^*$; $4=$DR. Only one of the contributions D, SR, SR$^*$ or
  DR can be calculated at a time.
\item {\tt iborn}: integer $\in \{ 0,1\}$. Born process ($2\to 2$). 
\item {\tt itwo}: integer $\in \{ 0,1\}$. NLO twobody corrections ($2\to 2$).
\item {\tt ithree}: integer $\in \{ 0,1\}$. The $2\to 3$ contributions
  above the $y_s$ cut. 
\item {\tt isplit}: integer $\in \{ 0,1\}$. Initial state
  singularity for the virtual photon ($2\to 2$).
\item {\tt iqcut}: integer $\in \{ 0,1\}$. Insert a finite cut-off
  $y_s$ ({\tt cutmin}) into the integration on the virtual photon side
  for the $2\to 3$ NLO contributions ($=1$) or not ($=0$). Setting
  {\tt iqcut=1} is necessary for photoproduction.
\end{itemize}
In the following the main settings for LO and NLO cross sections are
summarized in tabular form. These settings hold for all 4 values of
{\tt isdr}.
\begin{center}
\begin{tabular}[b]{lcccc} \hline
 & LO & NLO (DIS) & NLO ($Q^2=0$) & $\gamma^*\to q\bar{q}$ \\ \hline
{\tt iborn} & 1 & 1 & 1 & 0 \\
{\tt itwo}  & 0 & 1 & 1 & 0 \\
{\tt ithree}& 0 & 1 & 1 & 0 \\
{\tt isplit}& 0 & 0 & 1 & 1 \\
{\tt iqcut} & 0 & 0 & 1 & 0 \\ 
\hline
\end{tabular}
\end{center}

\subsubsection*{Initial State}

\begin{itemize}
\item {\tt Ea, Eb}: real*8. Energies of the incoming lepton a and
 hadron/lepton b in GeV. 
\item {\tt iwwa}: integer $\in \{0,1\}$. Selects formula for the photon flux
 in the case of photoproduction. For {\tt iwwa=0} the
 cross section is intergated analytically up to $Q^2_{max}$, whereas for
 {\tt iwwa=1} it is integrated using the value of {\tt thmax} below. 
\item {\tt thmax}: real*8. Maximum scattering angle of electron in 
 photoproduction for {\tt iwwa=1}.
\item {\tt P2max}: real*8 $\ge 0$. Maximum value of $P^2$
  ($\gamma^*\gamma$ mode).
\item {\tt iwwb}: integer $\in \{0,1\}$. see {\tt iwwa}
 ($\gamma^*\gamma$ mode).
\item {\tt thetbmax}: real*8. Maximum scattering angle of electron
 for {\tt iwwb=1} ($\gamma^*\gamma$ mode).
\end{itemize}

\subsubsection*{Subprocess}

\begin{itemize}
\item {\tt Nf}: real*8 $\in \{ 1,2,3,4,5\}$. Number of active
  flavours. 
\item {\tt lambda}: real*8 $>0$ in GeV. Value of $\Lambda_{QCD}$ for
  {\tt Nf} flavours. 
\item {\tt ialphas}: integer $\in \{ 1,2\}$. One-loop ({\tt
    ialphas}$=1$) or two-loop ({\tt ialphas}$=2$) formula for the 
  strong coupling constant $\alpha_s$ without thresholds.  For ${\tt
    ialphas}=3$ the value of $\alpha_s$ is  taken from the {\tt PDFLIB}
  (automatically adjusts $\Lambda$).
\item {\tt ycut}: real*8 $>0$. Value
  of $y_s$. The independence of the NLO cross sections on $y_s$ has
  been tested for $y_s\in [10^{-2},10^{-5}]$. Large statistical errors
  occur for too small $y_s$. Recommended value: $y_s=0.5\cdot10^{-3}$. 
\item {\tt idisga}: integer$\in \{0,1\}$. Since all partonic cross
  sections in \jv\ are implemented in the
  $\overline{\mbox{MS}}$-scheme, a photon PDF constructed in the
  DIS$_\g$ scheme has to be transformed into the
  $\overline{\mbox{MS}}$-scheme. This is done for the photon on side
  a by setting {\tt idisga}$=1$ \cite{4}. 
\item {\tt ipdftyp}: integer $\in \{ 1,2,3,4\}$. Selects the PDF for the
  resolved virtual photon. \\ $1=$ {\tt SaS} \cite{19}, $2=$ {\tt GRS}
  \cite{grs}, $3=$ {\tt PDFLIB} \cite{pdflib}
  (for real photons with $Q^2=0$).
\item {\tt igroupa}: integer. For {\tt ipdftyp}$=3,4$ this selects the
  group from the {\tt PDFLIB} (see manual \cite{pdflib}). For 
  {\tt ipdftyp}$=1$ this represents {\tt isasset}$\in \{
  1,2,3,4\}$, which selects input scale and scheme of SaS virtual
  photon PDF (see \cite{19}).
\item {\tt iseta}: integer. For {\tt ipdftyp}$=3,4$ this selects the
  set from the {\tt PDFLIB} (see manual \cite{pdflib}). For 
  {\tt ipdftyp}$=1$ this represents {\tt isasp2}$\in \{
  1,2,3,4,5,6,7\}$, which selects the scheme used to evaluate
  off-shell anomalous component in SaS virtual photon PDF (see \cite{19}).
\item {\tt igroupb}: integer. Selects the group from {\tt PDFLIB} for the
  resolved component for particle b (see manual \cite{pdflib}).
\item {\tt isetb}: integer. Selects the set from {\tt PDFLIB} (see
  manual \cite{pdflib}). 
\item {\tt a, b, c}: real*8 $>0$. Choosing the overall-scale $\mu^2$
  according to $\mu^2=a + bQ^2 + cp_T^2$. The $p_T$ is the largest 
  $E_T^{jet}$ of an event.
\end{itemize}

\subsubsection*{Phase space integration}

The phase space of \jv\ is generated in the following variables.
\begin{itemize}
\item {\tt Q2min, Q2max}: real*8 $\ge 0$. Minimum and maximum value of
  the photo virtuality $Q^2$. Setting ${\tt Q2min}=$0.d0 will produce a
  photoproduction cross  section.
\item {\tt ymin, ymax}: real*8 $\in \{ 0,1\}$. Minimum and maximum
  value of $y$.
\item {\tt ptmin}: real*8 $>0$. Minimum value of $E_T$ on parton
level. Set to $\frac12 E_{T,min}^{jet}$.
\end{itemize}

\subsubsection*{{\tt Vegas} and Output}

\begin{itemize}
\item {\tt ipoint}: integer $>0$. Defines the number of events
generated in the phase space defined above. Typical value: ${\tt
  ipoint}=10.000.000$. In NLO for smaller $y_s$ larger
  compensations occur and therefore the statistical errors become larger. 
\item {\tt itt}: integer $>0$. Number of iterations for {\tt Vegas}. 
  Recommended value: {\tt itt}$=5$.
\item {\tt jfileout}: character*20. Name of the output-file in which the
  histograms are stored.  
\end{itemize}

\section{Comparison with other programs}

\subsection{Infrared safe dijet cross sections}

Recently, a comparison of NLO programs for the DIS region has become
available \cite{comp}, which shows a good agreement between the
programs {\tt DISENT} and {\tt DISASTER}, systematic deviations of
{\tt MEPJET}, being typically of 5--8\% lower than the other programs
and finally the overall results of \jv\ being comparable to {\tt DISENT}
and {\tt DISASTER}. However, in some cases \jv\ deviates from these
programs by up to 20\%. Furthermore, a significant dependence of the
cross sections produced by \jv\ on the slicing parameter $y_s$ has been
reported. In the following, these results are discussed in detail and
the numbers for most scenarios described in \cite{comp} are
recalculated. 

The details of the technical settings for the comparison can be found
in \cite{comp}. Basically, dijet cross sections have been calculated
under HERA conditions in the Breit frame, using the inclusive
$k_\perp$ algorithm \cite{kt}. The authors define as a central
scenario the cuts
\begin{equation}
	30 < Q^2 < 40~\mbox{GeV}^2, \qquad 0.2 < y < 0.6 , \qquad
	E_T^{1,2} > 5~\mbox{GeV}
\end{equation}
where $E_T^1$ and $E_T^2$ are the highest and second highest $E_T$ of
an event. Then a number of additional cuts are proposed in six
scenarios, which are meant to ensure IR safeness of the jet
cross sections. In the following we will concentrate on the comparison
for scenarios 1--4, since scenarios 5 and 6 are not problematic and
sufficient agreement between the different NLO programs has been
found. As we will show later, the scenario 1 contains IR sensitive
cuts. In this section we concentrate on the IR safe scenarios, namely
scenarios 2--4, which are collected for reasons of completeness in the
Tab.~\ref{tab1}.

\begin{table}[hhh]
\begin{center}
\begin{tabular}[b]{r|r} \hline
No. & $E_{T1{\rm min}} / \GeV$\\ \hline
2 a) &    8 \\
2 b) &   15 \\
2 c) &  25 \\
2 d) &  40 \\
\hline
\end{tabular}
\hskip8mm
\begin{tabular}[b]{r|r|r}\hline
No. & $Q^2_{\rm min} / \GeV^2$ &  $Q^2_{\rm max} / \GeV^2$ \\ \hline
3 a) &    3  &   4 \\
3 b) &   30  &   40 \\
3 c) &  300  &  400 \\
3 d) & 3000  & 4000  \\
\hline
\end{tabular}
\hskip8mm
\begin{tabular}[b]{r|l|l}\hline
No. & $y_{\rm min}$ & $y_{\rm max}$ \\ \hline
4 a) &    0.01 & 0.05 \\
4 b) &   0.2 & 0.6 \\
4 c) &  0.9 & 0.95 \\
     & \\
\hline
\end{tabular}
\end{center}
\caption{\label{tab1}\it Different IR safe dijet scenarios. The $y$ and
$Q^2$ regions refer to an asymmetric cutting scenario with 
$E_{T1{\rm min}} = 8~\GeV$}
\end{table}

Recalculating the scenarios 2--4 with the new \jv\ version 2.1 using 
higher statistics and the optimized phase space generator we find the
results shown in appendix A, which mostly confirm the numbers from
\cite{comp}. Some numbers however show better agreement. \jv\ agrees
with {\tt DISENT} and {\tt DISASTER} on the 1\% level within the 
statistical errors. Only for scenario 2d \jv\ lies 2\% below
the {\tt DISENT}/{\tt DISASTER} results. This is a statistical problem
of the rather extreme scenario 2d. The cut on the higher $E_T$ jet is at
$40$~GeV, however all the $E_T$'s down to $2.5$~GeV have to be
generated on the parton level, since they can contribute to the cross
section after the jet recombination. Due to the steep fall-off of
the $E_T$ spectrum, too few events are generated at large
$E_T$. Overall, the agreement found between the  
three programs  {\tt DISENT}, {\tt DISASTER} and \jv\ is
very good, which is especially promising since the programs are based
on different methods to treat soft and collinear singularities. 

A reason for the systematic deviation of {\tt MEPJET} from the other three
programs, being overall about 5--8\% too small, might be due to 
not partially fractioned matrix elements \cite{mirkes}. In principle
the partial fractioning is not necessary if one goes to extremely
small $s_{min}$ values. This, however, leads to large statistical
errors and the calculation of precise results becomes
impractical. Problems of this kind in connection with the phase space
slicing technique are well known already from older works on jet
production from $e^+e^-$ anihilation (see e.g.\ \cite{gks84}).

\begin{table}[b]
\begin{center}
\begin{tabular}[b]{c|l|l}\hline
$y_s$ & scenario 2a & scenario 3d \\ \hline
$10^{-3}$ & 121.5\err 0.8 & 1.99 \err 0.02 \\
$10^{-4}$ & 121.4\err 2.2 & 2.0 \err 0.05 \\
$10^{-5}$ & 116.1\err 6.5 & 1.80 \err 0.13 \\ \hline
{\tt DISENT}& 119.5\err 0.3 & 1.985\err 0.003 \\
{\tt DISASTER}& 119.8\err 0.4 & 1.998\err 0.003 \\ \hline
\end{tabular}
\end{center}
\caption{\label{ys}\it  The dependence of \jv\ cross sections on the phase
space slicing parameter $y_s$ for two scenarios.}
\end{table}

A last point that has to be sorted out is the observed $y_s$
dependence of the \jv\ cross sections. Except for the scenarios 1b and
1c, which are discussed below to be IR sensitive, these dependences
are a statistical problem. We have observed that the adaptation
procedure implemented in the Monte-Carlo integration routine {\tt
VEGAS} \cite{vegas} becomes unreliable in the region of
small slicing parameters $y_s$ and the errors are drastically
underestimated. These kind of problems have been noticed recently and
an alternative integration algorithm has been proposed
\cite{metro}. To avoid such problems we have switched off the
adaptation in {\tt VEGAS} and simply generated an extremely large
number of events.  With this setting we have recalculated the $y_s$
dependence  for three values of the slicing parameter,
$y_s=10^{-3},10^{-4}$ and $10^{-5}$ for  scenarios 2a and 3d which are
among the critical cases in \cite{comp}. The results are shown in
Tab.~\ref{ys}. No dependence on $y_s$ within the statistical errors
is observed. Of course, the errors become larger, the 
smaller $y_s$ gets and it is very hard to calculate the cross section
for values of $y_s\le 10^{-5}$ since the compensation between large
positive and negative contributions lead to large statistical erros,
which is a well-known problem of the phase-space slicing
method. However, one can obtain reliable results with \jv\ for values
of $y_s$ between $10^{-3}$ and $10^{-4}$ and in this region also use
the adaptation technique implemented in {\tt VEGAS}. 

\subsection{Infrared sensitive scenarios}

The authors in \cite{comp} have proposed several ways to avoid IR
sensitive regions for the dijet cross sections. These are the
scenarios 1a--c, collected in Tab.~\ref{tab3}.

\begin{table}[hhh]
\begin{center}
\begin{tabular}[b]{r|c} \hline
No. & additional jet cut \\ \hline
1 a) &  $E_{T1{\rm min}} > 8~\GeV$ \\
1 b) &  $M_{jj} > 25~\GeV$  \\
1 c) &  $(E_{T1} + E_{T2}) > 17~\GeV$ \\ \hline
\end{tabular}
\end{center}
\caption{\label{tab3}\it Proposed ways to avoid IR sensitive regions.}
\end{table}

\begin{figure}[b]
  \unitlength1mm
  \begin{picture}(122,65)
  \put(10,-40){\epsfig{file=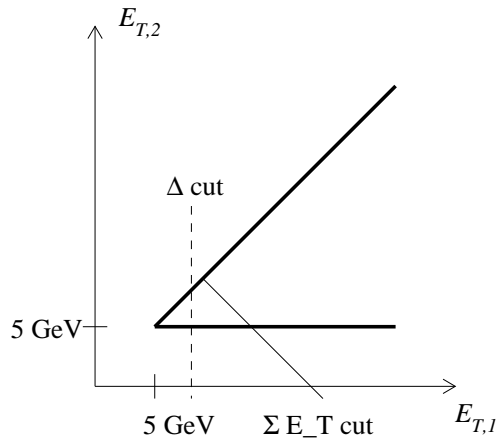,width=13cm}}
  \end{picture}
  \caption{\it The available phase space for the two highest $E_T$ jets.}
\end{figure}
Unfortunately, the scenarios 1b and 1c are {\it not} IR
safe. This can be understood in terms of the $E_T^1, E_T^2$ plane,
shown in Fig.~1. The thick solid lines show the symmetric cuts
$E_T^{1,2}>5$~GeV, which are known to be IR sensitive, due to the
point $E_T^1=E_T^2=5$~GeV \cite{5,ir}. The (negative) twobody contribution,
containing the analytic NLO corrections can not be compensated by
(positive) threebody contributions due to the limited phase space for
the third particle. To avoid this region, one can introduce 
a $\Delta$-cut on the highest $E_T$ jet,
$E_T^1=E_{T,min}+\Delta$ with $\Delta \simgt 1$~GeV. Furthermore, 
the proposed scenario 1c is shown in Fig.~1. However, the point 
$E_T^1=E_T^2=8.5$~GeV corresponding to a symmetric scenario is not
avoided by the $\sum E_T$ cut and not enough threebody
phase space is added to this point.

\begin{table}[t]
\begin{center}
\begin{tabular}{c|c|c|c|c} \hline
$\Delta$&  DISASTER++ & DISENT & JETVIP & MEPJET \\ \hline
0.0 (LO)  & 35.2 & 35.2 & 35.2 &  35.1 \\ 
0.0 (NLO) & 72.3 & 72.1 & 77.2 &  67.6 \\
0.125 & & 76.4 & 79.6 & \\
0.250 & & 76.6 & 79.7 & \\
0.375 & & 76.1 & 79.2 & \\
0.500 & & 75.6 & 78.1 & \\
0.625 & & 73.7 & 77.1 & \\
0.750 & & 72.3 & 75.5 & \\ 
2.000 & & 54.2 & 55.2 & \\ \hline
\end{tabular}
\end{center}
\caption{\label{della}\it The $\sum E_T$ scenario with an additional cut
$E_{T_1}=8.5~\GeV+\Delta$}
\end{table}
This IR sensitivity is the origin of the $y_s$ dependence of \jv. 
Furthermore, the cross sections obtained with two
different methods, namely the phase space slicing method and the
subtraction method, do not agree. Apart from the $y_s$ dependence of
\jv, an explicit way to see the IR sensitivity is to introduce a
$\Delta$-cut in addition to the $\sum E_T$ cut, i.e., to 
demand an additional cut on the higher $E_T$ jet with
$E_T^1=E_{T,min}+\Delta$~GeV, where $E_{T,min}=8.5$~GeV. We have
introduced such a cut and calculated the cross sections for various
$\Delta$ values for the two programs {\tt DISENT} and \jv. The results
are shown in Tab.~\ref{della}. Physically meaningful cross sections
will drop for rising $\Delta$. However, as can be seen in Tab.~\ref{della},
the NLO cross sections for the $\sum E_T$ scenario still rise up to $\Delta
=0.375$ and only reach the original value of the cross section around
$\Delta =0.625$. This behaviour is seen for both programs. We have checked
several values of $\Delta$ and found $\Delta =2$~GeV sufficient to
avoid IR sensitivity. As can be seen from the table, at this value
the programs \jv\ and {\tt DISENT} agree at the 2\% level. 

The same is true for scenario 1b, where
$M_{jj}>25$~GeV. This cut washes out the IR sensitive region, but
does not completely avoid it. This can again be seen in the $y_s$
dependence of the cross sections from \jv. By introducing an
additional $\Delta$-cut, as for the $\sum E_T$ case, one would again
see a rise of the NLO cross section with increasing $\Delta$.

\subsection{Photoproduction}

\begin{figure}[hhh]
  \unitlength1mm
  \begin{picture}(122,60)
    \put(2,-61){\epsfig{file=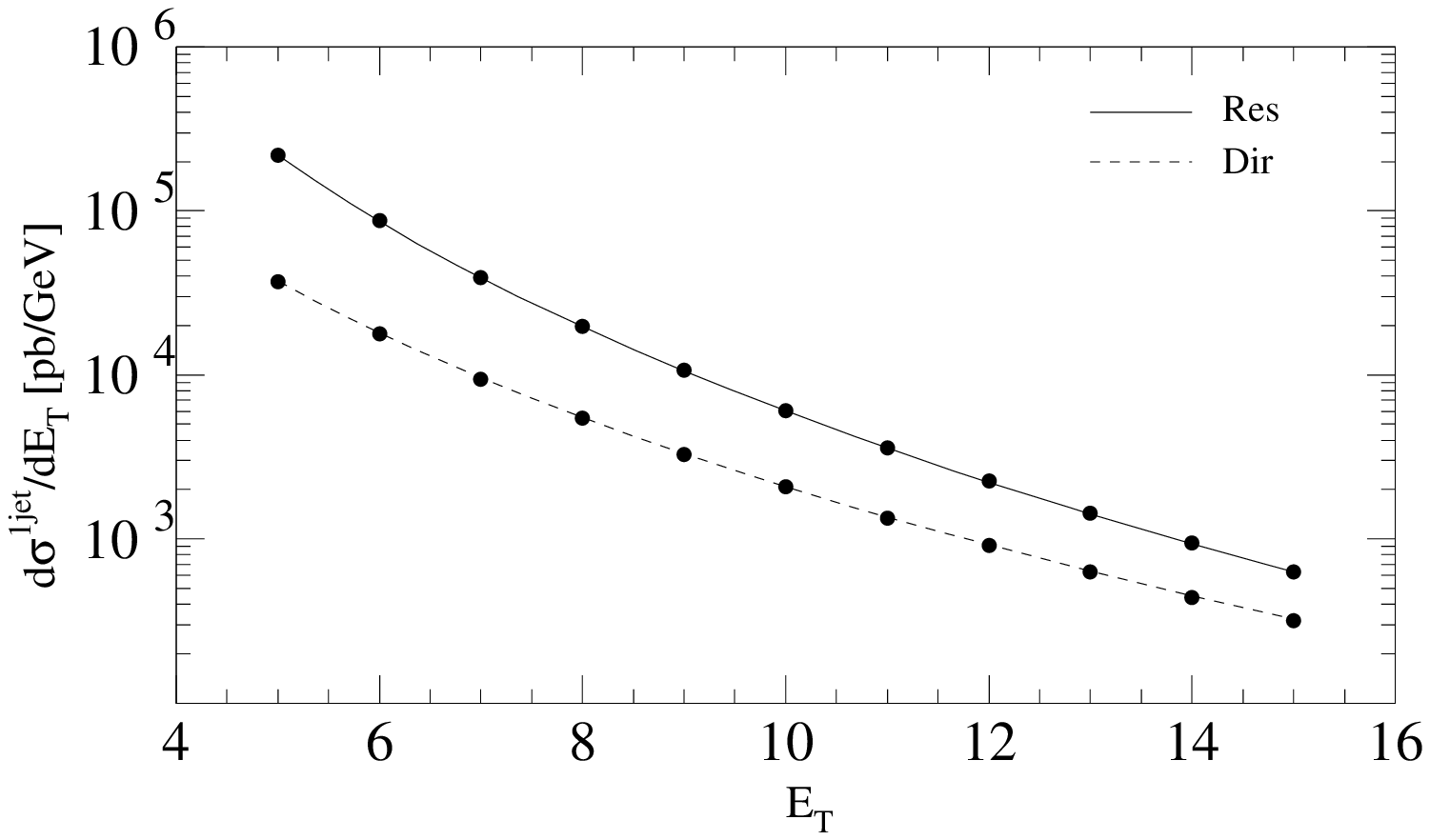,width=8.5cm,height=13cm}}
    \put(80,-61){\epsfig{file=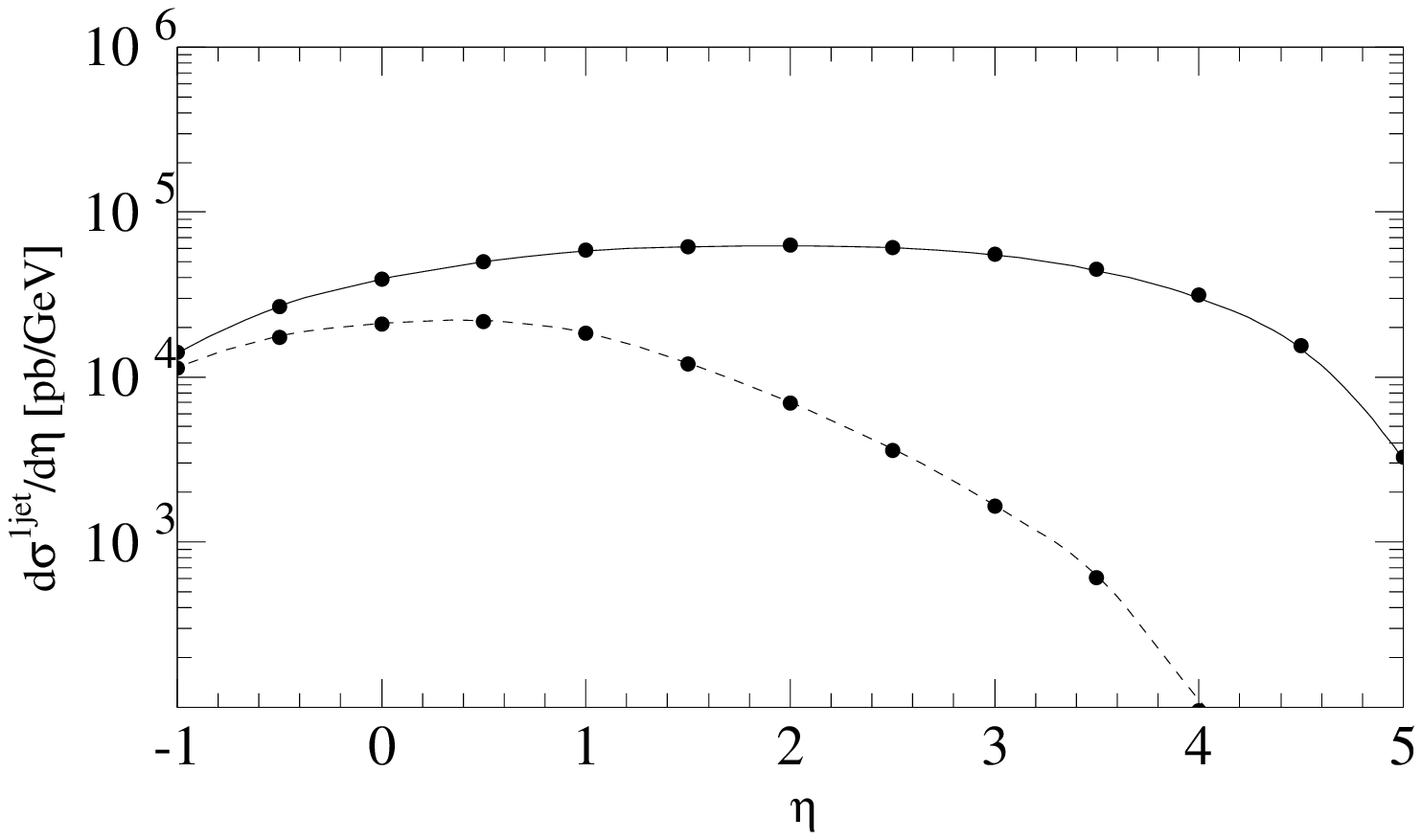,width=8.5cm,height=13cm}}
  \end{picture}
  \caption{\label{foto}\it Single-jet inclusive direct (lower curves)
  and resolved (upper curves) jet cross sections with \jv\  compared to
  the calculations \cite{Klasen:1996it} (dots) for photoproduction.}
\end{figure}

Recently, a comparison of calculations in jet photoproduction has been
performed \cite{hkv} in which the authors find that the three calculations
\cite{Frixione:1997ks,Harris:1997hz,Klasen:1996it} agree within the
statistical accuracy of the Monte Carlo integration. Only in certain
restricted regions of phase space, the calculations differ by up to
5\%. Since the program \jv\ allows to perform the limit
$Q^2\to 0$ starting from the DIS matrix elements, a comparison with
one of the photoproduction calculations will serve as a further
consistency check of the two methods for extracting collinear and soft
phase-space regions. We have therefore calculated inclusive
single-jet cross sections for photoproduction with \jv\ and compared
them to the calculations in \cite{Klasen:1996it}. The results are
shown in Fig.~\ref{foto} for the $E_T$ and $\eta$ spectra of the
direct (dashed) and resolved (full) components in NLO. The agreement
with the predictions from \cite{Klasen:1996it} (full dots) is perfect
for both components with differences below 1\%.  All other comparisons
performed so far between the two programs show a similar good agreement. 

Since the matrix elements for the resolved component are taken from
\cite{Klasen:1996it}, it is not surprising that it agrees
for both programs. The direct component in \jv\ is, however, 
completely independent from the calculations in \cite{Klasen:1996it}. The
excellent agreement of the two programs therefore establishes the consistency 
between the calculations for photoproduction and for the
DIS case for the phase space slicing method. Since the
equivalence of the subtraction and the  phase space slicing method has
been shown in \cite{hkv}, we conclude that \jv\ also agrees with the
calculations based on the subtraction method in the photoproduction
regime \cite{Frixione:1997ks}.

\subsection{Dependence on azimuthal angle}

We have implemented the full azimuthal dependence of the
jet cross sections in the new \jv\ program in LO according to the
formula 
\begin{equation} \label{phi}
	\frac{d\sigma}{d\phi} = A + B\cos\phi + C\cos 2\phi \ , 
\end{equation}
where $\phi$ denotes the azimuthal angle of the jets around the
virtual photon direction in the hadronic cms frame. The lepton plane
defines $\phi =0$. The coefficients $A,B,C$ are related to the
polarization density elements of the virtual photon. 

\begin{figure}[b]
  \unitlength1mm
  \begin{picture}(122,105)
    \put(30,-4){\epsfig{file=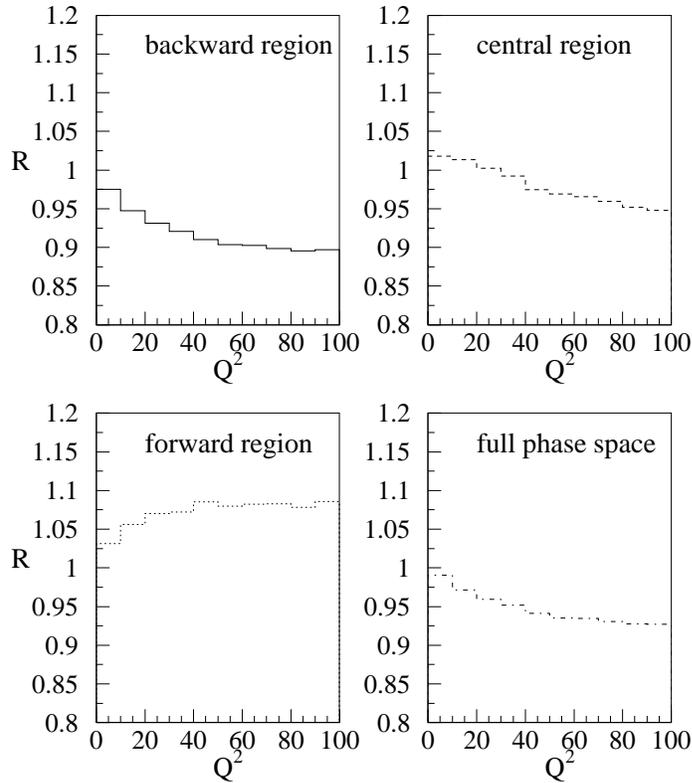,width=10cm}}
  \end{picture}
  \caption{\label{phib}\it The ratio of unintegrated over integrated LO
  cross sections in four different rapidity regions in the lab frame.}
\end{figure}

The dependence in (\ref{phi}) can be integrated out if no cuts on the
jets are imposed in the laboratory frame. This is, however, seldom the
case experimentally. In the following we check how large the effect of
using the integrated and unintegrated matrix elements is in LO.
We calculate cross sections in LO within the central scenario defined
above with no additional cuts and probe the region $Q^2\in
[0,100]$~GeV$^2$. In Fig.~\ref{phib} three rapidity regions are shown
in the laboratory frame, namely
\begin{equation} 
 -1 < \eta_{lab} < 0.5 , \qquad 0.5 < \eta_{lab} < 1.5 , \qquad 1.5 <
  \eta_{lab} < 2.8 ,
\end{equation}
corresponding to the backward, central and forward region of the
detector. We have plotted the ratio of the integrated over the unintegrated
matrix elements, denoted as $R$. As one sees, for $Q^2<10$~GeV$^2$ the
$\phi$-dependence is below 2.5\% for all regions and therefore
negligible. In the central region the dependence stays below 5\%,
whereas in the more extreme regions the dependence can reach 10\% for
the largest $Q^2$-bin shown. This is in accordance with the results in
\cite{mw}. We have further checked that the difference between
integrated and unintegrated has a maximum at $Q^2=100$~GeV$^2$. For
larger virtualities, the difference again diminuishes and falls below
5\% for all rapidity regions above $Q^2=500$~GeV$^2$.
If the NLO corrections stay reasonably small, e.g., around
20\%, one can estimate the error due to cuts in the laboratory-frame
in NLO to be below 2\%, for any value of $Q^2$. 

We conclude that for the main region of applicability of \jv, namely
the region of $Q^2<10$~GeV$^2$, where also the resolved cross sections
plays a role, the $\phi$-dependence is described correctly in the
present \jv\ version. For all other $Q^2$, the dominant contribution
has been taken into account and a correct description is given if the
NLO corrections are not too large. In any case, the
corrections will not be larger than 10\%.

\section{Summary}

We have presented an updated version of the \jv\ 1.1 program for
jet production in $eP$- and $\gamma^*\gamma$-scattering. The most
important changes are:
\begin{itemize}
\item a full event record is provided on the parton level
\item the complete analysis of the events has to be done in the {\tt
      user} routine
\item the phase space is generated in $(y,Q^2)$-space
\item the $\phi$-dependence has been implemented in LO
\end{itemize}

We have numerically shown the equivalence of two methods used to
extract singular phase space regions, namely, the subtraction method and the
phase-space slicing method, by comparing 4 existing NLO programs. 
The programs {\tt DISENT}, {\tt DISASTER} and {\tt JetViP} agree on
the 1\% level within the statistical errors of integration. Any
differences observed before in these programs are either of
statistical nature or due to IR sensitive cutting scenarios. In
particular, we have demonstrated the IR sensitivity of the $\sum E_T$
scenario. The numerical equivalence of phase space slicing and subtraction
method have furthermore been underpinned by comparisons in the
photoproduction regime. 

The difference between using azimuthal dependent matrix elements in LO and
those where this behaviour has been integrated out is negligible for
all rapidity regions for $Q^2<10$~GeV$^2$. By taking into account the
azimuthal dependence  of the matrix elements in LO also cross sections
involving cuts in the laboratory frame are described correctly in the 
present \jv\ version for all practical purposes.

\subsection*{Acknowledgments}

I have profited from discussions with T.~Carli, C.~Duprel, D.~Graudenz,
G.~Grindhammer, Th.~Hadig, G.~Kramer, T.~Sch\"orner, M.H.~Seymour and
M.~Wobisch. I thank T.~Sch\"orner for providing me the {\tt DISENT}
numbers for Tab.~\ref{della}.


\newpage

\begin{appendix}

\section{Numerical results}

Results of the comparison between the programs {\tt DISENT, DISASTER,
MEPJET} and \jv\ in the DIS region for IR safe scenarios 2--4. The
numbers are taken from \cite{comp}, except for the NLO \jv\ numbers, which
have been recalculated.

\renewcommand{\arraystretch}{1.3}
\begin{center}
\begin{tabular}{c|r|r|r|r} \hline
scenario  &  DISASTER++ & DISENT & JETVIP  & MEPJET \\ \hline \hline
2 a) & 119.8 \err 0.4 &  119.5 \err 0.3 & 121.5 \err 0.8 & 113.5 \err 0.2 \\
LO:&41.66 \err 0.08 &  41.77 \err 0.06 &  41.75 \err 0.03 & 41.72 \err 0.03 \\
\hline
2 b) & 16.58 \err 0.09 & 16.53 \err 0.05 & 16.21 \err 0.07 & 15.74 \err 0.08 \\
LO:&6.19 \err 0.02 & 6.22 \err 0.01 & 6.214 \err 0.005 & 6.221 \err 0.003 \\
\hline
2 c) & 2.08 \err 0.03 & 2.052 \err 0.008 & 2.04 \err 0.1 & 1.908 \err 0.008 \\
LO: & 1.023 \err 0.005 & 1.022 \err 0.002 & 1.025 \err 0.001 & 1.025
\err 0.0005 \\ \hline
2 d) & 0.140 \err 0.005 & 0.140 \err 0.001 & 0.122 \err 0.015 & 0.123
 \err 0.005 \\ 
 LO: & 0.120 \err 0.001 & 0.1213 \err 0.0004 & 0.1207 \err 0.0002 &
 0.1209 \err   0.00006    \\ \hline
\end{tabular}

\bigskip

\begin{tabular}{c|r|r|r|r} \hline
scenario  &  DISASTER++ & DISENT & JETVIP & MEPJET \\ \hline \hline
3 a) & 341.2 \err 1.7 & 339.1 \err 1.2 & 340.1 \err 0.7  &  331.5 \err 0.4 \\
LO: & 48.4 \err 0.1 & 48.42 \err 0.08 & 48.36 \err 0.04 & 48.40 \err 0.04 \\ 
\hline 
3 b) & as 2 a) & & & \\ \hline
3 c) &26.85 \err 0.06 & 26.68 \err 0.05 & 26.72 \err 0.09 & 24.68 \err 0.05 \\ 
LO: & 16.94 \err 0.02 & 16.93 \err 0.02 & 16.93 \err 0.01 & 16.92 \err 0.01 \\ 
\hline
3 d) & 1.998 \err 0.003 & 1.985 \err 0.003 & 1.995 \err 0.007 & 1.8917 \err 0.0038 \\
LO: & 1.498 \err 0.002 & 1.497 \err 0.001 & 1.496 \err 0.001 & 1.497
\err 0.001 \\ \hline
\end{tabular}

\bigskip

\begin{tabular}{c|r|r|r|r} \hline
scenario  &  DISASTER++ & DISENT & JETVIP & MEPJET \\ \hline \hline
4 a) & 19.2 \err 0.1 & 18.96 \err 0.07 & 18.67 \err 0.04 & 17.19 \err 0.04  \\
LO:  & 11.61 \err 0.04 & 11.57 \err 0.02 & 11.59 \err 0.01 &
 11.587 \err 0.006 \\  \hline
4 b) &  as 2 a)   &    &    &     \\ \hline
4 c) & 6.42 \err 0.03 & 6.45 \err 0.02 & 6.40 \err 0.02  & 6.09 \err 0.03  \\
LO:  & 2.161 \err 0.006 & 2.162 \err 0.003 & 2.17 \err 0.01 &
 2.160 \err  0.002    \\ \hline
\end{tabular}
\end{center}

\newpage

\section{Example of steering file}

The following steering file is for calculating the NLO direct cross
section under HERA conditions in DIS with scales $\mu^2=Q^2$. 

\begin{footnotesize}
\begin{tt}
\begin{tabbing}
'========================================================================='\\
'	\qquad\qquad \=       $\!\!\!\!\!\!\!\!\!$ CONTRIBUTION'\= \\
'========================================================================='\\
1       \>        iproc   [1=ep; 2=ee] \\
2       \>        isdr    [1=D; 2=SR; 3=SR*; 4=DR]\\
1       \>        iborn   [2->2 Born]\\
1       \>        itwo    [2->2 singular contributions (for NLO)]\\
1       \>        ithree  [2->3 contribution]\\
0       \>        isplit  [gamma->qq term]\\
0       \>        iqcut   [yqi-min in 2->3 matrices? 1=yes; 0=no]\\
'========================================================================='\\
'	\>	INITIAL STATE'\\
'========================================================================='\\
27.5d0	\>	Ea	[energy of lepton a] \\
820.d0	\>	Eb	[energy of proton/ lepton b] \\
'------- Lepton a --------------------------------------------------------'\\
0       \>        iwwa  [which Weizs.-Will.: 0=ln(Q2mx/Q2mn); 1=ln(thmax/me)]\\
180.d0  \>        thmax [max angle]\\
'------- Lepton b (only relevant for the ee-case ) -----------------------'\\
4.d0    \>      P2max   [P2=virtuality of real photon]\\
0       \>      iwwb    [which Weizs.-Will.: 0=ln(P2mx/P2mn); 1=ln(thmax/me)]\\
180.d0  \>      thetbmx [max angle for Weizs.-Will]\\
'========================================================================='\\
'        \>       SUBPROCESS'\\
'========================================================================='\\
5.d0     \>       Nf      [Number of active flavours]\\
0.204d0  \>       lambda  [Lambda-QCD (has to match Nf)]\\
3        \>       ialphas [QCD coupling: 1=one-loop; 2=two-loop; 3=PDFLIB]\\
1d-3     \>       y-cut   [phase-space-slicing parameter]\\
'------ PDFs for the resolved contributions ------------------------------'\\
0        \>       idisga  [DISg -> MSbar for photon a]\\
1        \>       ipdftyp [PDF for y*(res): 1=SaS;2=GRS;3=DG;4=PDFLIB] \\
2        \>       igroupa (Param. for SaS or PDFLIB --> see manual)    \\
2        \>       iseta   ( " )                         \\               
0        \>       idisgb  [DISg -> MSbar for photon b]\\
4        \>       igroupb [authors of PDF on side b: y(res) or prot]\\
34       \>       isetb   [Set-No]\\
'------- Scales ----------------------------------------------------------'\\
0.d0    \>       a       [Scale: mu**2=a+b*Q**2+c*pt**2]\\
1.d0     \>       b  \\     
0.d0     \>       c\\
\end{tabbing}
\end{tt}\end{footnotesize}

\newpage

\begin{footnotesize}
\begin{tt}
\begin{tabbing}
'========================================================================='\\
'	\qquad\qquad \=       $\!\!\!\!\!\!\!\!\!$ PHASE SPACE INTEGRATION'\= \\
'========================================================================='\\
5d0      \>       Q2min   [0.d0 selects photoproduction]\\
10d0     \>       Q2max   \\
0.1d0    \>       ymin    [min y]\\
0.6d0    \>       ymax    [max y]\\
2.25d0   \>       ptmin   [ptmin for parton in hadr cms; =Etj-min/2]\\
'========================================================================='\\
'        \>       VEGAS and OUTPUT'\\
'========================================================================='\\
10000000 \>       ipoin   [no of events produced in phase space above]\\
5         \>      itt     [no of iterations]\\
'nlo.out'  \>  jfileout [Filename] \\
\end{tabbing}
\end{tt}\end{footnotesize}

\end{appendix}
\end{document}